\documentclass[12pt]{article}

\usepackage[utf8]{inputenc}
\usepackage[utf8]{inputenc}

\usepackage{amssymb}
\usepackage{amsfonts}
\usepackage{graphicx}
\usepackage{circuitikz}
\usetikzlibrary{quantikz}

\usepackage[verbose]{placeins} 

\usepackage{braket}
\usepackage{tikz}
\usepackage[T1]{fontenc}
\usepackage{imakeidx}

\usepackage{xcolor,varwidth}

\usepackage{pgfplots}
 \usetikzlibrary{backgrounds,calc}
 \usepackage{amsmath,tikz}
 \usetikzlibrary{quotes,angles}
\pgfplotsset{compat = newest}
\usetikzlibrary{quotes,angles}

\usetikzlibrary{svg.path}
\usepackage{pgfornament}

\usepackage{xcolor}

\usepackage{subfig}
\usepackage{graphicx}
\def\mset #1[#2]=#3{%
	\expandafter\xdef\csname #1#2\endcsname{#3}
}
\def\mget #1[#2]{%
	\csname #1#2\endcsname
}
\def\minc #1[#2]+=#3{%
	\pgfmathparse{\mget #1[#2]+#3}%
	\mset #1[#2]=\pgfmathresult
}

\date{}

\usepackage{hyperref}
\usepackage [all]{hypcap}
\usepackage{color}

\usepackage{listings}
\definecolor{codegreen}{rgb}{0,0.6,0}
\definecolor{codegray}{rgb}{0.5,0.5,0.5}
\definecolor{codepurple}{rgb}{0.58,0,0.82}
\definecolor{backcolour}{rgb}{0.95,0.95,0.92}

\lstdefinestyle{mystyle}{
    backgroundcolor=\color{backcolour},   
    commentstyle=\color{codegreen},
    keywordstyle=\color{magenta},
    numberstyle=\tiny\color{codegray},
    stringstyle=\color{codepurple},
    basicstyle=\ttfamily\footnotesize,
    breakatwhitespace=false,         
    breaklines=true,                 
    captionpos=b,                    
    keepspaces=true,                 
    numbers=left,                    
    numbersep=5pt,                  
    showspaces=false,                
    showstringspaces=false,
    showtabs=false,                  
    tabsize=2
}

\begin{document}

\title{Quantum Entanglement, Quantum Teleportation, Multilinear Polynomials and Geometry  }
\author{ Juan M. Romero \thanks{jromero@cua.uam.mx }, Emiliano Montoya-Gonz\'alez \thanks{emiliano.montoya.g@cua.uam.mx } and  Oscar Velazquez-Alvarado  \thanks{oscar.velazquez@cua.uam.mx }\\
Departamento de Matemáticas Aplicadas y Sistemas,\\
Universidad Aut\'onoma Metropolitana-Cuajimalpa,\\
M\'exico, D.F 05300, M\'exico }
\date{\today}

\maketitle
\begin{abstract}
We show that quantum entanglement states are associated with  multilinear polynomials that cannot be factored. 
By using these multilinear polynomials, we  propose a  geometric representation for entanglement states. In particular, we show that the Bell's states are associated with non-factorable real multilinear polynomial, which can be represented geometrically by  three-dimensional surfaces.  Furthermore,  in this  framework, we show that a quantum circuit can be seen as a geometric transformations of plane geometry. This phenomenon is analogous to gravity, where  matter curves space-time.   In addition, we show an analogy between  quantum teleportation and operations involving  multilinear polynomials. 
\end{abstract}

\section{\label{sec:introduction}Introduction}
Quantum entanglement and quantum  teleportation are both important phenomena in quantum physics \cite{yano,Lee,bernt,Nishioka}. 
In particular, quantum computing exploits these phenomena  to create  diverse  algorithms, including those for  quantum cryptographic  \cite{yano,Lee,bernt,Nishioka}.  It is worth mentioning that recently, quantum computing  has proven to be superior to classical computing. For example, the  Fast Fourier Transform (FFT) is considered  one of the most important algorithms in classical computing, which has computational complexity of $O(n2^{n})$. However,  the Quantum  Fourier Transform   has $O(n^{2})$  computational complexity \cite{Asaka}.  Thus, the  Quantum  Fourier Transform is more efficient than  the classical FFT, a review about quantum computational complexity can be seen in \cite{Wastrous}.  For these reasons, quantum computing development will have a significant impact on simulation to understand various physics systems in high energy \cite{he}, nuclear physics \cite{nuclear}, quantum chemistry \cite{chemistry}, etc. In addition, in quantum cryptography, when  the  Shor's algorithm could be implemented error-free, it would break many classical public-key cryptography schemes  \cite{Lee}. For these reasons several  industries have begun to explore the benefits of quantum computing \cite{Herman}. \\ 

Now, quantum  algorithms can be represented graphically as an  array of interconnected quantum gates, where the quantum gates are unitary matrices. These array of quantum gates are called quantum circuits.  Now, by quantifying the minimal size of any circuit that implements a given unitary, a  complexity can be associated to a quantum circuit, the so called quantum circuit complexity. Notably,  a geometry can be obtained from quantum circuit complexity \cite{Bengtsson}. In this regard, an interesting result is that  some of theses geometries can be related with systems such as black-holes \cite{Brown,Brown1,Haferkamp,Chapman} .\\

Thus,  quantum computing is not only  essential for improving the efficiency of simulations but also  important  for advancing our knowledge of fundamental physics.\\

In this paper we show that quantum entanglement states and quantum teleportation have an analogy with multilinear polynomials and their geometry.
In particular, we show that quantum entanglement states are equivalent to bilinear polynomials  that cannot be factored as two independent polynomials. 
Note that the coefficients of associated  multilinear polynomials are real or complex numbers, then these polynomials can be represented in real or complex geometry.
In this respect,  we show that  the Bell's states can be related with bilinear polynomials and  with   surfaces in $\mathbb{R}^{3}.$ In addition, we show that
 the initial states in  quantum circuits are  associated with a plane geometry, then a quantum circuit  can be seen as geometric transformations of plane geometry.  This phenomenon is similar to gravity, where  matter curves space-time. Furthermore, we show an analogy between  quantum teleportation andoperations involving  multilinear polynomials. . \\

This paper is organized as follows.  In Sec. \ref{Poly} we show that entanglement states are associated with bilinear polynomials. In Sec. \ref{Bell} we show that Bell's states are associated with real bilinear polynomials and surfaces in $\mathbb{R}^{3}.$  In Sec. \ref{Tel} we show a case of   quantum teleportation is associated with operations in multilinear polynomials.   In Sec. \ref{Tel1} we show that quantum teleportation is associated with operations in multilinear polynomials.  In Sec. \ref{Con} a summary is given. 

\section{Entanglement states, bilinear polynomial and geometry }
\label{Poly}
Now, suppose that we have  a system with two qubits, then if
\begin{eqnarray}
c_{1}, c_{2}, c_{3},c_{4}\label{c1}
\end{eqnarray}
are complex constants which satisfy 
\begin{eqnarray}
|c_{1}|^{2}+| c_{2}|^{2}+|c_{3}|^{2}+|c_{4}|^{2}=1\nonumber
\end{eqnarray}
we can construct the state 
\begin{eqnarray}
\ket{\psi}=c_{1}\ket{00}+c_{2}\ket{01}
+c_{3}\ket{10}+c_{4}\ket{11}.\label{c1A1}
\end{eqnarray}
As well, we can construct the  two states
\begin{eqnarray}
\ket{\psi_{1}}&=&\gamma_{1}\ket{0}+\gamma_{2}\ket{1},\nonumber\\
\ket{\psi_{2}}&=&\lambda_{1}\ket{0}+\lambda_{2}\ket{1},\nonumber
\end{eqnarray}
and the state
\begin{eqnarray}
\ket{\psi_{1}}\otimes\ket{\psi_{2}}=\gamma_{1}\lambda_{1}\ket{00}+\gamma_{1}\lambda_{2}\ket{01}
+\gamma_{2}\lambda_{1}\ket{10}+\gamma_{2}\lambda_{2}\ket{11}.\nonumber
\end{eqnarray}
Then, if the state $\ket{\psi}$ is separable,  there are  two states $\ket{\psi_1},\ket{\psi_2}$ such that 
\begin{eqnarray}
\ket{\psi}=\ket{\psi_{1}}\otimes \ket{\psi_{2}}.
\end{eqnarray}
Notice that in this case  the  equations 
\begin{eqnarray}
c_{1}&=&\gamma_{1}\lambda_{1}, \label{con1}\\
c_{2}&=&\gamma_{1}\lambda_{2},\label{con2}\\
c_{3}&=&\gamma_{2}\lambda_{1},\label{con3} \\
c_{4}&=& \gamma_{2}\lambda_{2} \label{con4}
\end{eqnarray}
are satisfied.  In addition,  if we define the matrix
\begin{eqnarray}
A=\begin{pmatrix}
c_{1}&c_{2}\\
c_{3}& c_{4}
\end{pmatrix}=\begin{pmatrix}
\gamma_{1}\lambda_{1}&\gamma_{1}\lambda_{2}\\
\gamma_{2}\lambda_{1}& \gamma_{2}\lambda_{2}
\end{pmatrix}
\nonumber
\end{eqnarray}
we have
\begin{eqnarray}
\det A=\gamma_{1}\lambda_{1}\gamma_{2}\lambda_{2}-\gamma_{2}\lambda_{1}\gamma_{1}\lambda_{2}=0.
\nonumber
\end{eqnarray}
Namely, if $\det A=0,$ the state $\ket{\psi}$ is a not separable state. In other words, if $\det A\not =0,$ then $\ket{\psi}$ is an entangled state. Then, for each one matrix of  $L(2,\mathbb{C})$ we can obtain an entangled state.\\

In addition, by using the constants  \eqref{c1},  the following   bilinear polynomial
\begin{eqnarray}
 P(x,y)=c_{1}+c_{2}x+c_{3}y+c_{4}xy  \label{bl1}
\end{eqnarray}
can be constructed as well. Observe that if the  bilinear polynomial \eqref{bl1} can be factorized with two linear function 
\begin{eqnarray}
Q_{1}(x)&=&\gamma_{1}+\gamma_{2}x,\nonumber\\
Q_{2}(y)&=&\lambda_{1}+\lambda_{2}y,\nonumber
\end{eqnarray}
then  the  equation
\begin{eqnarray}
P(x,y)&=&c_{1}+c_{2}x+c_{3}y+c_{4}xy\nonumber\\
&= & Q_{1}(x)Q_{2}(y)=(\gamma_{1}+\gamma_{2}x)(\lambda_{1}+\lambda_{2}y)\nonumber \\
&=&\gamma_{1}\lambda_{1}+\gamma_{1}\lambda_{2}y
+\gamma_{2}\lambda_{1}x+\gamma_{2}\lambda_{2}xy\nonumber
\end{eqnarray}
must be satisfied. This last  equation implies the equations \eqref{con1}-\eqref{con4}. Hence, if $\det A=0$ the bilinear polynomials $P(x,y)$ is factorizable. Namely, 
 if $\det A\not =0$ the bilinear polynomials $P(x,y)$ is not factorizable. Therefore, an entanglement states of two qubits $\ket{\psi}$ \eqref{c1A1} has associated a not factorizable bilinear polynomial $P(x,y)$ \eqref{bl1}. \\

Notice that the bilinear polynomials are associated with a geometry and because the constant \eqref{c1} are complex number, the geometry is complex as well. In particular,  if these constants \eqref{c1}
are real numbers,  the multilinear polynomials $P(x,y)$ represent a surface in $\mathbb{R}^{3}.$ Then, in this last case the  entanglement states of two qubits $\ket{\psi}$ have associated a  surface in $\mathbb{R}^{3}.$\\

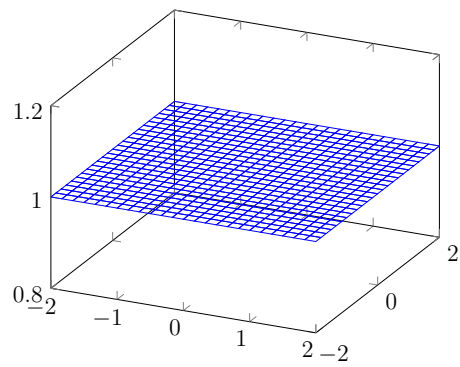
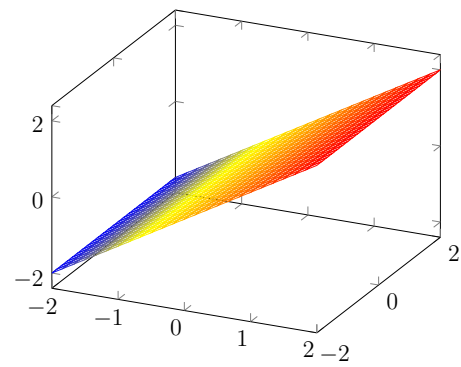
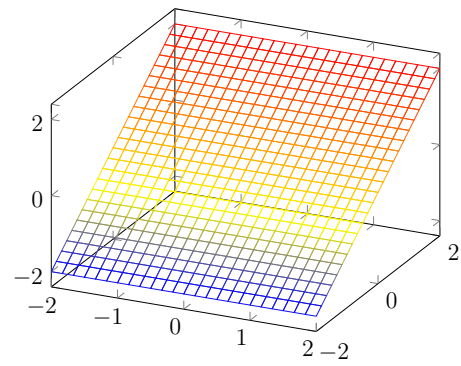
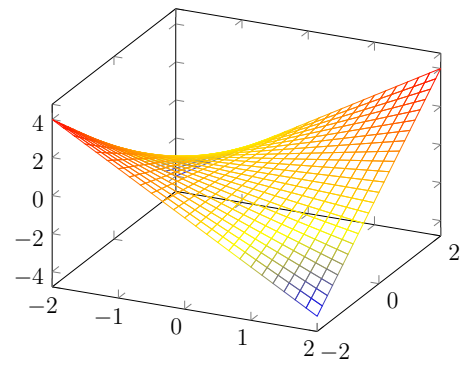
\begin{figure}
     \centering
     \subfloat[]{ \begin{tikzpicture}[scale=0.750]
    \begin{axis}[]
        \addplot3 [mesh,domain=-2:2] {1};
    \end{axis}
\end{tikzpicture} }      \subfloat[]{ \begin{tikzpicture}[scale=0.750]
    \begin{axis}[]
        \addplot3 [mesh,domain=-2:2] {x};
    \end{axis}
\end{tikzpicture} } \\ 
   \subfloat[]{ \begin{tikzpicture}[scale=0.750]
    \begin{axis}[]
        \addplot3 [mesh,domain=-2:2] {y};
    \end{axis}
\end{tikzpicture} }     
 \subfloat[]{ \begin{tikzpicture}[scale=0.750]
    \begin{axis}[]
        \addplot3 [mesh,domain=-2:2] {x*y};
    \end{axis}
\end{tikzpicture} }     \caption{ (a) Polynomial $f_{1}(x,y)=1$.  (b) Polynomial $f_{2}(x,y)=x.$ (c) Polynomial $f_{3}(x,y)=y.$ (d)  Polynomial $f_{4}(x,y)=xy$. }
   \label{fig:rep-pita}
\end{figure}

For example, we can see that the bilinear polynomial \eqref{bl1} is a linear combination of the functions
\begin{eqnarray}
f_{1}(x,y)&=&1,\label{f1}\\
f_{2}(x,y)&=&x,\label{f2}\\
f_{3}(x,y)&=&y,\label{f3}\\
f_{4}(x,y)&=&xy.\label{f4}
\end{eqnarray}
These function are associated with the surfaces showed in the figure \eqref{fig:rep-pita}.

\section{Bell's states}
\label{Bell}

Notice that by using the matrices  
\begin{eqnarray}
B_{1}=\frac{1}{\sqrt{2}}\begin{pmatrix}
1&0\\
0&1
\end{pmatrix},  B_{2}=\frac{1}{\sqrt{2}}\begin{pmatrix}
1&0\\
0&-1
\end{pmatrix},  \\B_{3}=\frac{1}{\sqrt{2}}\begin{pmatrix}
0&1\\
1&0
\end{pmatrix},  B_{4}=\frac{1}{\sqrt{2}}\begin{pmatrix}
0&1\\
-1&0
\end{pmatrix}, \nonumber \label{ma1}
\end{eqnarray}
we have the Bell's states 
\begin{eqnarray}
\ket{B_{1}}&=&\frac{1}{\sqrt{2}}\left(\ket{00}+\ket{11}\right),\label{bell1} \\
\ket{B_{2}}&=&\frac{1}{\sqrt{2}}\left(\ket{00}-\ket{11}\right),\label{bell2} \\
\ket{B_{3}}&=&\frac{1}{\sqrt{2}}\left(\ket{01}+\ket{10}\right),\label{bell3} \\
\ket{B_{4}}&=&\frac{1}{\sqrt{2}}\left(\ket{01}-\ket{10}\right). \label{bell4} 
\end{eqnarray}
where we can obtain
\begin{eqnarray}
\ket{00}&=&\frac{1}{\sqrt{2}}\left(\ket{B_1}+\ket{B_2}\right),\label{bas1}\\
\ket{11}&=&\frac{1}{\sqrt{2}}\left(\ket{B_1}-\ket{B_2}\right),\label{bas2}\\
\ket{01}&=&\frac{1}{\sqrt{2}}\left(\ket{B_3}+\ket{B_4}\right),\label{bas3}\\
\ket{10}&=&\frac{1}{\sqrt{2}}\left(\ket{B_3}-\ket{B_4}\right). \label{bas4}
\end{eqnarray}

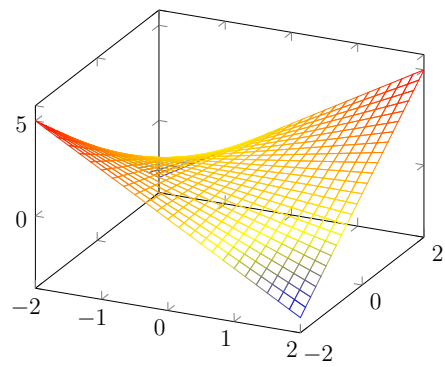
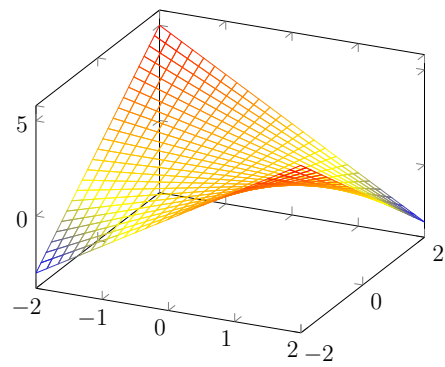
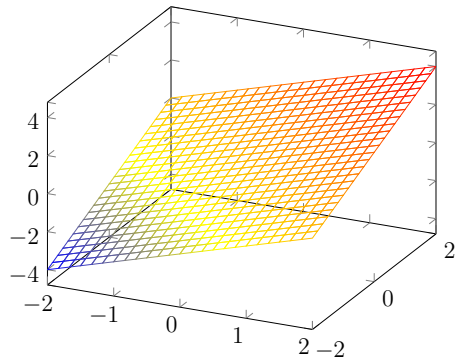
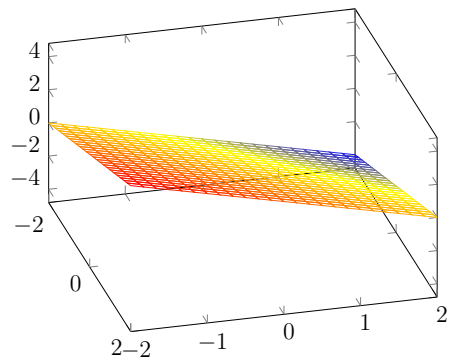
\begin{figure}
     \centering
     \subfloat[]{ \begin{tikzpicture}[scale=0.750]
    \begin{axis}[]
        \addplot3 [mesh,domain=-2:2] {1+y*x};
    \end{axis}
\end{tikzpicture}}     
 \subfloat[]{ \begin{tikzpicture}[scale=0.750]
    \begin{axis}[]
        \addplot3 [mesh,domain=-2:2] {1-x*y};
    \end{axis}
\end{tikzpicture} } \\ 
   \subfloat[]{ \begin{tikzpicture}[scale=0.750]
    \begin{axis}[]
        \addplot3 [mesh,domain=-2:2] {x+y};
    \end{axis}
\end{tikzpicture} }     
 \subfloat[]{ \begin{tikzpicture}[scale=0.750]
    \begin{axis}[view={75}{40}]
        \addplot3 [mesh,domain=-2:2] {x-y};
    \end{axis}
\end{tikzpicture} } 
      \caption{ (a) Polynomial $P_{1}(x,y)=1+xy$.  (b) Polynomial $P_{2}(x,y)=1-xy.$ (c) Polynomial $P_{3}(x,y)=x+y.$  (d) Polynomial $P_{4}(x,y)=x-y$. }
   \label{fig:rep-pita2}
\end{figure}

Also, by using the matrices \eqref{ma1} we can obtain the no factorizable bilinear polynomials
\begin{eqnarray}
P_{1}(x,y)&=&\frac{1}{\sqrt{2}}\left(1+xy\right) ,\nonumber \\
P_{2}(x,y)&=&\frac{1}{\sqrt{2}}\left(1-xy\right),\nonumber \\
P_{3}(x,y)&=&\frac{1}{\sqrt{2}}\left(x+y\right),\nonumber\\
P_{4}(x,y)&=&\frac{1}{\sqrt{2}}\left(x-y\right). \nonumber
\end{eqnarray}
which implies the equations
\begin{eqnarray}
1&=&\frac{1}{\sqrt{2}}\left(P_1(x,y)+P_2(x,y)\right),\label{bp1} \\
xy&=&\frac{1}{\sqrt{2}}\left(P_1(x,y)-P_2(x,y)\right),\label{bp2}\\
x&=&\frac{1}{\sqrt{2}}\left(P_3(x,y) +P_4(x,y)\right),\label{bp3}\\
y&=&\frac{1}{\sqrt{2}}\left(P_3(x,y) -P_4(x,y)\right). \label{bp4}
\end{eqnarray}
Additionally, the bilinear polynomials are associated with the surfaces which are presented in the figure  \ref{fig:rep-pita2}.
Then, for each one of Bell's states  there are an associated bilinear polynomial and a surface. \\

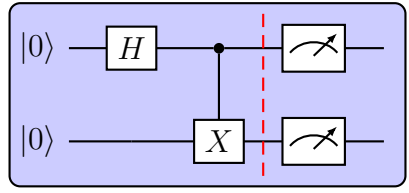
\begin{figure}[hbt!]
\centering
\begin{quantikz}%[color=black,background color=yellow]
\gategroup[wires=2,steps=6,style={rounded corners,fill=blue!20}, background]{}
&\lstick{$|{0}\rangle$} & \gate{H}&\ctrl{1} & \meter{} & \qw
\\
&\lstick{$|{0}\rangle$}  & \qw & \gate{X} \slice{} & \meter{} & \qw
\end{quantikz}
\caption{Quantum circuit for the first Bell's state \eqref{bell1}}
    \label{fig:BellCirc1}
\end{figure}

 Now, in the Figure \ref{fig:BellCirc1} is given the quantum circuit to obtain  the first Bell's state \eqref{bell1} and their qiskit code is given in Listing \ref{lst:CBell1}. Notice that in the  quantum circuit 
the initial  state is $\ket{00},$ which is associated with the plane given in the figure \ref{fig:rep-pita}(a). In addition, the   final  state is  \eqref{bell1}, which is associated with the surface given in the figure \ref{fig:rep-pita2}(a). 
Then,   the quantum circuit can be seen as geometric transformation that transforms the plane geometry in the surface \ref{fig:rep-pita2}(a). 
Furthermore, due that in  all quantum circuit 
the initial  state is $\ket{00\cdots 0}$, which is associated with a plane geometry, all quantum circuit can be  seen as a geometric transformations of plane geometry.  Observe that this phenomenon is analogous to gravity, where  matter curves space-time.  

\begin{lstlisting}[language=Python, caption=qiskit code to obtain  the first Bell's state \eqref{bell1},  label={lst:CBell1} ]
import qiskit as q
# Create two quantum registers
qr = q.QuantumRegister(2, 'q')
# Create two classic registers
cr = q.ClassicalRegister(2, 'c')
# Create a circuit with the four registers
circuit = q.QuantumCircuit(qr, cr)
# Add the gates to the circuit
circuit.h(qr[0])
circuit.cx(qr[0], qr[1])
# Measure the two qubits
circuit.measure(qr, cr)
\end{lstlisting}

\section{Quantum teleportation and multilinear Polynomial, an example}
\label{Tel}

Now, let us remember some issues  about quantum  teleportation.
First, consider a state
\begin{eqnarray}
\ket{\phi}=\gamma_{1}\ket{0}+\gamma_{2}\ket{1}. \label{tpq-sp}
\end{eqnarray}
Then we can construt the state
\begin{eqnarray}
\ket{\phi}\ket{B_{1}}=\frac{1}{\sqrt{2}}(\gamma_{1}\ket{00}\ket{0}+\gamma_{1}\ket{01}\ket{1}\\ +\gamma_{2}\ket{10}\ket{0}+\gamma_{2}\ket{11}\ket{1})\nonumber
\end{eqnarray}
which can be written as
\begin{eqnarray}
\ket{\phi}\ket{B_{1}}=\frac{1}{\sqrt{2}}(\gamma_{1}\ket{00}\ket{0}+\gamma_{1}\ket{01}\ket{1}\\ +\gamma_{2}\ket{10}\ket{0}+\gamma_{2}\ket{11}\ket{1})\nonumber
\end{eqnarray}
Furthermore, by using the states \eqref{bas1}-\eqref{bas4} we obtain
obtain
\begin{eqnarray}
\ket{\phi}\ket{B_{1}}&=&\frac{1}{2}\ket{B_1}(\gamma_{1}\ket{0}+\gamma_{2}\ket{1})\nonumber \\
&& +\frac{1}{2}\ket{B_2}(\gamma_{1}\ket{0}-\gamma_{2}\ket{1})
\nonumber \\
&& +\frac{1}{2}\ket{B_3}(\gamma_{2}\ket{0}+\gamma_{1}\ket{1})\nonumber \\
&& +\frac{1}{2}\ket{B_4}(-\gamma_{2}\ket{0}+\gamma_{1}\ket{1}). \label{qtb1}
\end{eqnarray}

Now, we will study the equivalent operation by using  multilinear polynomial. First, we consider the function
\begin{eqnarray}
Q(z)=\gamma_{1}+\gamma_{2}z, \label{pol-pt}
\end{eqnarray}
then we obtain
\begin{eqnarray}
Q(z)P_{1}(x,y)&=&(\gamma_{1}+\gamma_{2}z)\frac{1}{\sqrt{2}}\left(1+xy\right) \nonumber \\
&=&\frac{1}{\sqrt{2}}\left(\gamma_{1}+\gamma_{1}xy +\gamma_{2}z+\gamma_{2}zxy\right) \nonumber \\
&=&\frac{1}{\sqrt{2}}\left(( \gamma_{1}+\gamma_{2}z)+(\gamma_{1}y +\gamma_{2}zy)x\right). \label{qtp}
\end{eqnarray}
Now, using the variables $(y,z)$ instead the variables $(x,y)$ in the equations \eqref{bp1}-\eqref{bp4} we obtain
\begin{eqnarray}
1&=&\frac{1}{\sqrt{2}}\left(P_1(y,z)+P_2(y,z)\right),\nonumber \\
yz&=&\frac{1}{\sqrt{2}}\left(P_1(y,z)-P_2(y,z)\right),\nonumber\\
y&=&\frac{1}{\sqrt{2}}\left(P_3(y,z) +P_4(y,z)\right),\nonumber\\
z&=&\frac{1}{\sqrt{2}}\left(P_3(y,z) -P_4(y,z)\right). \nonumber
\end{eqnarray}

Then, by substitution  these last equations into the expression \eqref{qtp} we arrive to 
\begin{eqnarray}
Q(z)P_{1}(x,y)&=&\frac{1}{\sqrt{2}}\left(( \gamma_{1}+\gamma_{2}z)+(\gamma_{1}y +\gamma_{2}zy)x\right) \nonumber \\
&=&\frac{1}{2}\Big(( \gamma_{1}\left(P_1(y,z)+P_2(y,z)\right)\nonumber \\
&& +\gamma_{2}\left(P_3(y,z) -P_4(y,z)\right))\nonumber\\ 
& &+(\gamma_{1}\left(P_3(y,z) +P_4(y,z)\right) \nonumber \\
&& +\gamma_{2}\left(P_1(y,z) -P_2(y,z)\right))x\Big) \nonumber \\
&=&\frac{1}{2}\Big( P_1(y,z)( \gamma_{1}+\gamma_{2}x) \nonumber \\
&& +  P_2(y,z)( \gamma_{1}-\gamma_{2}x)+\nonumber \\& &+ P_3(y,z) ( \gamma_{2}+\gamma_{1}x)\nonumber \\ 
&& + P_4(y,z) (- \gamma_{2}+\gamma_{1}x)
\Big) \nonumber 
\end{eqnarray}
namely
\begin{eqnarray}
Q(z)P_{1}(x,y)
&=&\frac{1}{2}\Big( P_1(y,z)( \gamma_{1}+\gamma_{2}x) \nonumber \\
&& +  P_2(y,z)( \gamma_{1}-\gamma_{2}x)+\nonumber \\& &+ P_3(y,z) ( \gamma_{2}+\gamma_{1}x)\nonumber \\
&& + P_4(y,z) (- \gamma_{2}+\gamma_{1}x)
\Big). \nonumber 
\end{eqnarray}
This expression is equivalent to the equation \eqref{qtb1}

\section{Quantum  teleportation and multilinear polynomial, general case}
\label{Tel1}

Now, in order to study  the general case we renamed the states
\begin{eqnarray}
\ket{C_{1}}&=&\ket{00},\nonumber  \\
\ket{C_{2}}&=&\ket{01},\nonumber \\
\ket{C_{3}}&=&\ket{10},\nonumber \\
\ket{C_{4}}&=&\ket{11}. \nonumber
\end{eqnarray}
 Then, we can propose the states 
\begin{eqnarray}
\ket{V_{1}}&=&a_{11}\ket{C_1}+a_{12}\ket{C_2}  +a_{13}\ket{C_3}  +a_{14}\ket{C_4},      \nonumber \\
\ket{V_{2}}&=&a_{21}\ket{C_1}+a_{22}\ket{C_2}  +a_{23}\ket{C_3}  +a_{24}\ket{C_4} ,     \nonumber \\
\ket{V_{3}}&=&a_{31}\ket{C_1}+a_{32}\ket{C_2}  +a_{33}\ket{C_3}  +a_{34}\ket{C_4},      \nonumber \\
\ket{V_{4}}&=&a_{41}\ket{C_1}+a_{42}\ket{C_3}  +a_{43}\ket{C_3}  +a_{44}\ket{C_4}.      \nonumber 
\end{eqnarray}
From these states the following matrices 
\begin{eqnarray}
M_{1}=\begin{pmatrix}
a_{11} & a_{12}\\
a_{13}& a_{14}
\end{pmatrix},
 M_{2}=\begin{pmatrix}
a_{21} & a_{22}\\
a_{23}& a_{24}
\end{pmatrix}, \\ \nonumber
M_{3}=\begin{pmatrix}
a_{31} & a_{32}\\
a_{33}& a_{34}
\end{pmatrix},
M_{4}=\begin{pmatrix}
a_{41} & a_{42}\\
a_{43}& a_{44}
\end{pmatrix}
\end{eqnarray}
can be defined. Then if the equations 
\begin{eqnarray}
\det M_{i}=
a_{i1}a_{i4}- a_{i2}a_{i3}\not =0,\quad i=1,2,3,4
 \nonumber
\end{eqnarray}
are satisfied,  the states $\ket{V_{i}}$ are entangled.  \\

In addition, notice that by using the matrix
\begin{eqnarray}
T=
\begin{pmatrix}
a_{11}&a_{12}&a_{13}&a_{14}\\
a_{21}&a_{22}&a_{23}&a_{24}\\
a_{31}&a_{32}&a_{33}&a_{34}\\
a_{41}&a_{42}&a_{43}&a_{44}\\
\end{pmatrix}\label{maT}
\end{eqnarray}
 the states $\ket{V_{i}}$  can be written as
\begin{eqnarray}
\ket{V_{i}}=T_{ij}\ket{C_{j}}. \nonumber
\end{eqnarray}

Moreover, if   $\ket{V_{j}}$ are orthonormal states the following equation
\begin{eqnarray}
\braket{V_{i}|V_{j}}=\delta_{ij}.  \nonumber
\end{eqnarray}
must be satisfied. Notice, that this last equation can be written as 
\begin{eqnarray}
\braket{V_{i}|V_{j}}&=&a_{i1}^{*}a_{j1}+a_{i2}^{*}a_{j2}+a_{i3}^{*}a_{j3}+a_{i4}^{*}a_{j4}\nonumber\\
&=&a_{il}^{*}a_{jl}=a_{il}^{*}a_{lj}^{T}=\delta_{ij}\nonumber
\end{eqnarray}
namely
\begin{eqnarray}
a_{il}^{*}a_{lj}^{T}=\delta_{ij},\nonumber
\end{eqnarray}
which implies 
\begin{eqnarray}
TT^{*T}=I,\nonumber
\end{eqnarray}
that is
\begin{eqnarray}
T^{-1}=T^{\dagger}=T^{*T}.\nonumber
\end{eqnarray}
Thus,  $T$ is an unitary matrix
\begin{eqnarray}
T\in U(4). \nonumber
\end{eqnarray}

From this last result, we obtain
\begin{eqnarray}
\ket{C_j}=T_{jk}^{-1}\ket{V_{k}}=T_{jk}^{T*}\ket{V_{k}}=T_{kj}^{*}\ket{V_{k}},
\end{eqnarray}
which can be written as  
\begin{eqnarray}
\ket{C_j}=a_{kj}^{*}\ket{V_{k}}. \label{cb-in}
\end{eqnarray}

Then, by using the state \eqref{tpq-sp}, we have 
\begin{eqnarray}
\ket{\phi_{i}}&=&\ket{\psi}\otimes \ket{V_{i}}\nonumber\\
&=&\ket{00}\left(\gamma_{1}a_{i1}\ket{0}+\gamma_{1}a_{i2}\ket{1}\right)\nonumber \\
&& +\ket{01}\left( \gamma_{1}a_{i3}\ket{0}+\gamma_{1}a_{i4}\ket{1}\right)\nonumber\\
& &+\ket{10}\left(\gamma_{2}a_{i1}\ket{0}+\gamma_{2}a_{i2}\ket{1}\right)\nonumber \\
&& +\ket{11} \left(\gamma_{2}a_{i3}\ket{0}+\gamma_{2}a_{i4}\ket{1}\right) \nonumber
\end{eqnarray}
which can be written as
\begin{eqnarray}
\ket{\phi_{i}}&=&\ket{C_1}\left(\gamma_{1}a_{i1}\ket{0}+\gamma_{1}a_{i2}\ket{1}\right)\nonumber \\
&& +\ket{C_2}\left( \gamma_{1}a_{i3}\ket{0}+\gamma_{1}a_{i4}\ket{1}\right)\nonumber\\
& &+\ket{C_3}\left(\gamma_{2}a_{i1}\ket{0}+\gamma_{2}a_{i2}\ket{1}\right)\nonumber \\
&& +\ket{C_4} \left(\gamma_{2}a_{i3}\ket{0}+\gamma_{2}a_{i4}\ket{1}\right). \nonumber
\end{eqnarray}
Furthermore, using the equation \eqref{cb-in}  we obtain
\begin{eqnarray}
\ket{\phi_{i}}& =&\ket{V_k}\Bigg( \ket{0}( \gamma_{1}( a^{*}_{k1} a_{i1}+a^{*}_{k2} a_{i3})\nonumber \\
&& +\gamma_{2}(
a^{*}_{k3}a_{i1}+a^{*}_{k4}a_{i3}))\nonumber\\
& &+\ket{1}(\gamma_{1}( a^{*}_{k1} a_{i2}+
a^{*}_{k2}a_{i4})\nonumber \\
&& +\gamma_{2}(a^{*}_{k3}a_{i2}+
a^{*}_{k4}a_{i4}))\Bigg). \nonumber
\end{eqnarray}
This last expression can be written as 
\begin{eqnarray}
\ket{\phi_{i}}&=&\ket{V_{k}}\Big(\gamma^{\prime}_{1ki}\ket{0}+\gamma^{\prime}_{2ki}\ket{1} \Big), \label{qtbg1}
\end{eqnarray}
where
\begin{eqnarray}
\begin{pmatrix}
\gamma_{1ki}^{\prime}\\
\gamma_{2ki}^{\prime}
\end{pmatrix}
=\left(M_{k}^{*}M_{i}\right)^{T}
\begin{pmatrix}
\gamma_{1}\\
\gamma_{2}
\end{pmatrix}.\nonumber
\end{eqnarray}

Now, if a observer $O_{1}$ measure the state
$\ket{V_{k}}$
the observer $O_{2}$ have to applied the gate 
\begin{eqnarray}
\left(\left(M_{k}^{*}M_{i}\right)^{T}\right)^{-1}. \nonumber
\end{eqnarray}
It can be showed that if the matrices
\begin{eqnarray}
M_{i}
\end{eqnarray}
are unitary matrices, then the matrices 
\begin{eqnarray}
G=\left(M_{k}^{*}M_{i}\right)^{T}  \nonumber
\end{eqnarray}
are unitary. In this case we have the inverse
\begin{eqnarray}
G^{-1}=M_{k}M_{i}^{*}.  \nonumber
\end{eqnarray}

Now, we can propose the following multilinear polynomials
\begin{eqnarray}
R_{1}(x,y)&=&a_{11}+a_{12}x  +a_{13}y +a_{14}xy,      \nonumber \\
R_{2}(x,y)&=&a_{21}+a_{22}x  +a_{23}y +a_{24}xy,      \nonumber \\
R_{3}(x,y)&=&a_{31}+a_{32}x  +a_{33}y +a_{34}xy,      \nonumber \\
R_{4}(x,y)&=&a_{41}+a_{42}x  +a_{43}y +a_{44}xy.      \nonumber 
\end{eqnarray}
Now, we defined the bilinear polynomial. 
Then, by using the matrix \eqref{maT} and the functions \eqref{f1}-\eqref{f4}, we obtain 
\begin{eqnarray}
R_{i}(x,y)&=&a_{ij}f_{j}(x,y)=T_{ij}f_{j}(x,y),  \label{cb-pt1}\\
f_{j}(x,y)&=&T_{jk}^{-1}R_{k}(x,y)=a_{kj}^{*}R_{k}(x,y). \label{cb-pt2}
\end{eqnarray}

Now, by using the linear function (\ref{pol-pt}) we get
\begin{eqnarray}
\phi_{i}(x,y,z)&=&Q(z)R_{i}(x,y)\nonumber\\
&=&\left(\gamma_{1}a_{i1}+\gamma_{1}a_{i2}x\right)+y\left( \gamma_{1}a_{i3}+\gamma_{1}a_{i4}x\right)\nonumber \\
&& +z\left(\gamma_{2}a_{i1}+\gamma_{2}a_{i2}x\right)\nonumber \\& &+yz\left(\gamma_{2}a_{i3}+\gamma_{2}a_{i4}x\right) \nonumber
\end{eqnarray}
which  can be written as
\begin{eqnarray}
\phi_{i}(x,y,z)&=&Q(z)R_{i}(x,y)\nonumber\\
&=&f_{1}(y,z)\left(\gamma_{1}a_{i1}+\gamma_{1}a_{i2}x\right)\nonumber \\
&& +f_{2}(y,z)\left( \gamma_{1}a_{i3}+\gamma_{1}a_{i4}x\right)+ \nonumber\\
& &+f_{3}(y,z)\left(\gamma_{2}a_{i1}+\gamma_{2}a_{i2}x\right)\nonumber \\
&& +f_{4}(y,z)\left(\gamma_{2}a_{i3}+\gamma_{2}a_{i4}x\right). \nonumber
\end{eqnarray}
Furthermore, by using the equation \eqref{cb-pt2}  we obtain
\begin{eqnarray}
\phi_{i}(x,y,z)& =&R_{k}(y,z)\Bigg(( \gamma_{1}( a^{*}_{k1} a_{i1}+a^{*}_{k2} a_{i3})\nonumber \\
&& +\gamma_{2}(
a^{*}_{k3}a_{i1}+a^{*}_{k4}a_{i3}))+\nonumber\\
& &+x(\gamma_{1}( a^{*}_{k1} a_{i2}+
a^{*}_{k2}a_{i4})\nonumber \\
&& +\gamma_{2}(a^{*}_{k3}a_{i2}+
a^{*}_{k4}a_{i4}))\Bigg) .\nonumber
\end{eqnarray}
This last expression can be written as 
\begin{eqnarray}
\phi_{i}(x,y,z) =R_{k}(y,z)\Big(\gamma^{\prime}_{1ki}+\gamma^{\prime}_{2ki}x \Big),\nonumber
\end{eqnarray}
which is equivalent to the equation \eqref{qtbg1}
\section{Conclusions}
\label{Con}

We have demonstrated that quantum entanglement states are associated with multilinear polynomials that cannot be factored. Subsequently, by utilizing these multilinear polynomials, we have established a geometry associated with entanglement states.
In this respect,  
 we have shown that the Bell's states have associated non-factorable real multilinear polynomial  and with  three dimensional surfaces. 
Now, in general because in  all quantum circuits the initial  state and final state  are different, and the initial state $\ket{00\cdots 0}$ is associated with a plane geometry,  we have shown  that all quantum circuits can be  seen as a geometric transformations of plane geometry.
 Notice that
when the final state is associated with a curved geometry, this phenomenon is analogous to gravity, where  matter curves space-time. 
In addition, we demonstrated an analogy between quantum  teleportation  and operations involving multilinear polynomial. \\

We believe that the results of this paper provide  another framework to study the quantum circuit and their associated geometry. In a future work, from this point of view,  we will study quantum circuits associated  with curved spacetime geometry.\\

\end{document}